\documentclass[aps,pra,twocolumn,eqsecnum,amsmath,superscriptaddress,showpacs]{revtex4-1}
\usepackage{graphicx}
\usepackage{longtable}
\usepackage{epsfig}
\usepackage{dcolumn}
\usepackage{bm}
\usepackage{amssymb}
\usepackage{multirow}
\usepackage{times,color}
\usepackage{hyperref}

\begin{document}

\title{Electronic properties in a quantum well structure of Weyl semimetal
}

\author{Wen-Long You}
\affiliation{College of Physics, Optoelectronics and Energy, Soochow
University, Suzhou, Jiangsu 215006, People's Republic of China}

\author{Xue-Feng Wang}
\email{wxf@suda.edu.cn}
\affiliation{College of Physics, Optoelectronics and Energy, Soochow
University, Suzhou, Jiangsu 215006, People's Republic of China}
\affiliation{
Key Laboratory of Terahertz Solid-State Technology,
Shanghai Institute of Microsystem and Information Technology,
Chinese Academy of Sciences, 865 Changning Road, Shanghai 200050, China
}

\author{Andrzej M. Ole\'s}
\affiliation{Max-Planck-Institut f\"ur Festk\"orperforschung,
             Heisenbergstrasse 1, D-70569 Stuttgart, Germany }
\affiliation{Marian Smoluchowski Institute of Physics, Jagiellonian University,
             prof. S. \L{}ojasiewicza 11, PL-30348 Krak\'ow, Poland }

\author{Jiao-Jiao Zhou}
\affiliation{College of Physics, Optoelectronics and Energy, Soochow
University, Suzhou, Jiangsu 215006, People's Republic of China}

\date{\today}
\begin{abstract}
We investigate the confined states and transport of three-dimensional
Weyl electrons around a one-dimensional external rectangular
electrostatic potential. Confined states with finite transverse wave
vector exist at energies higher than the half well depth or lower than
the half barrier height. The rectangular
potential appears completely transparent to normal incident electrons
but not otherwise. The tunneling transmission coefficient is sensitive
to their incident angle and shows resonant peaks when their energy
coincides with the confined spectra. In addition, for electrons in the
conduction (valence) band through a potential barrier (well), the
transmission spectrum has a gap of width increasing with the incident
angle. Interestingly, the electron linear zero-temperature conductance
over the potential can approach zero when the Fermi energy is aligned
to the top and bottom energies of the potential, when only electron
beams normal to the potential interfaces can pass through. The
considered structure can be used to collimate Weyl electron beams.
\end{abstract}


\maketitle

In the standard picture a topologically-nontrivial phase of matter
corresponds to gapped bulk materials with topologically protected
gapless surface/edge states \cite{Kane10,Zhang11}. Recent work has
shown, however, that certain gapless systems may also be topologically
nontrivial which are known as topological semimetals
\cite{Bur11,Qi13,Wit14,Vis14,Gibson15,Turner13}. They arise from the
existence of band-touching points (Weyl nodes) in their electronic
structure. Their properties become particularly striking when the Fermi
energy approaches the Weyl nodes at which a linear energy dispersion in
three dimensions exists.

Weyl semimetal (WSM) is one of the topological semimetals which embeds
splitting Weyl nodes without other degeneracy by breaking time-reversal
symmetry or spatial inversion symmetry \cite{Bal11}.
Theoretically, a Weyl node can be modeled as a magnetic
monopole in momentum space which cannot exist independently.
Therefore, the Weyl nodes occur always in pairs with opposite chirality
and consequently the Fermi arc states emerge on the surface
\cite{Wan11,Bal11,Xu11,Baum15}.
Due to these unique characteristics, WSMs can show various exotic phenomena,
such as transport anomaly \cite{Aji12,Parameswaran14,Qi15}, high Chern
number quantum anomalous Hall effect states \cite{Jiang12,Skirlo15} and
electrical optical physics \cite{Xie15,Yang15}.
WSMs have been theoretically proposed in a number of candidate systems,
including Rn$_2$Ir$_2$O$_7$ pyrochlore \cite{Yan11}, zinc-blende
lattice \cite{Ojanen13}, ABi$_{1-x}$Sb$_x$Te$_3$ (A=La or Lu)
\cite{JPLiu14}, HgCr$_2$Se$_4$ \cite{Xu11}, TaAs
\cite{Weng15,Xu15}, TaP \cite{Shekhar15}.
Recent experiments have provided convincing evidences that TaAs and
NiP are indeed WSMs \cite{Xu15,Lv15,Zhang15,Shekhar-np2015}.

Beside the WSM, another type of topological semimetal named as Dirac
semimetal have also attracted considerable interest. The Dirac
semimetals have been experimentally confirmed in Na$_3$Bi
\cite{Liu14,Xu-Sci15} and Cd$_3$As$_2$ \cite{ZKLiu14}. The Weyl nodes
also exist in Dirac semimetals. Therefore, the Dirac semimetal often behaves like WSM.
For example, it is suggested by the transport properties of Dirac
semimetal material Cd$_3$As$_2$ as reported by several groups
\cite{Wan13,Liang15,YLiu14,Che15,Li15}.

However, there are two
Weyl fermions with opposite chiralities in each Weyl node of Dirac semimetals
due to the presence of both time reversal and inversion symmetry.
The non-degeneracy of the intersecting bands in WSMs warrants a topological stability of the Weyl nodes, which hold advantage
for possible practical applications in nanodevices. Such a unique electronic system has been a
particularly attractive platform for investigation of various electric
and optical properties. The appearance of Weyl nodes will generate
negative magnetoresistance related to chiral anomaly under the presence
of parallel magnetic and electric fields \cite{Nielsen83}. A number of
compelling observations have been made recently
\cite{XCHuang15,XJYang15,Shekhar15}.

Effective electric field or electrostatic potential can be established in binary/trinary materials using the band engineering methods.
Well developed techniques such as the molecular-beam epitaxy (MBE) have been widely employed in semiconductor industry to
fabricate perfect GaAs/Al$_x$Ga$_{1-x}$As quantum wells (QWs) for electronic devices. Similar technologies can be used
to build p-n-p or n-p-n junctions in WSMs and manipulate the properties of the systems.
A remarkable property of Weyl
electrons is their chiral behavior, which will lead to the absence of
their back-scattering and the corresponding difficulty of being
confined by electrostatic potentials.
In this work, we study how to confine three-dimensional (3D) Weyl electrons around
a one-dimensional (1D) QW or quantum barrier (QB) and
discuss the scattering properties of incident 3D Weyl electrons by the 1D quantum structure.
We will consider an infinite model system of electrons near one Weyl node.
In finite real materials with multiple nodes, we can sum up the contributions from all the nodes if the short range internode coupling
and the surface-state effect are negligible. This is valid for low-energy electrons in large high-quality samples
with wide and properly oriented QW.
Our analysis suggests that a planar transistor composed of WSM heterojunction QWs or QBs can be used to collimate normally incident electron beams.

Weyl electrons are spin-1/2 chiral fermions described by the
$2\times 2$ 3D Weyl equation. The Hamiltonian of the WSMs system is
given by
\begin{eqnarray}
\hat{H}=  v_F(\vec{\sigma}\cdot \vec{p}) + U(x) ,
\label{Ham1}
\end{eqnarray}
where the Pauli matrices form a vector,
$\vec{\sigma}=\{\sigma_x,\sigma_y,\sigma_z\}$ and $U(x)$ is the 1D
external potential. Here $v_F$ specifies the Fermi velocity, and for
simplicity we assume in the following $v_F$ $\sim$ $ 10^6$ m/s which
allows for a comparison with graphene \cite{Vas06}. Note that it might
be anisotropic and have different values along $v_x$, $v_y$ and $v_z$
in real materials such as Na$_3$Bi \cite{Liu14}.
However, the physics is unchanged by rescaling the corresponding scales
in momentum space.

We consider a 1D square QW $U(x)=-U_0 \theta(L/2-|x|)$ of depth $U_0=50$ meV and width $L=200$ nm.
The results for other values of $U_0$ and $L$ can be obtained by a
scaling method. If the 1D QW is replaced by a 1D QB of height $U_0$ and
width $L$, behaviors of electrons at energy $\epsilon$ is the same as
that of electrons at energy $-\epsilon$ in the QW. For the sake of generality,
some dimensionless parameters, $\xi = x/L$, $\beta = k_y L$, $\gamma = k_z L$,
$\epsilon =E L/\hbar v_F$, and $u = U(x) L/\hbar v_F$, will be used in the paper.
Due to the translational invariance in the $y$- and $z$-directions, the wave functions have the form
$\Psi_C(x,y,z)=\phi_C(x) e^{ik_y y} e^{ik_z z}$ with $C=A, B$
referring to their two spinor components.

For confined states, the spinor components decay exponentially outside
the potential well, $\vert \xi \vert > 1/2$, characterized by a decay
constant,
$\hat{\alpha}=-i\alpha$ for $\alpha^2\equiv\epsilon^2-\beta^2-\gamma^2<0$,
while they appear standing waves depicted by wave vector $\kappa$ with
$\kappa^2\equiv (\epsilon+u_0)^2-\beta^2 - \gamma^2 > 0$ inside the potential well.
Employing the boundary condition, analytical wave functions can be obtained and the energy eigenvalues are determined by the equation
$2 \hat{\alpha} \kappa \cot(\kappa) =\kappa^2-\hat{\alpha}^2-u_0^2.$

\begin{figure}[t!]
\includegraphics[width=8.2cm]{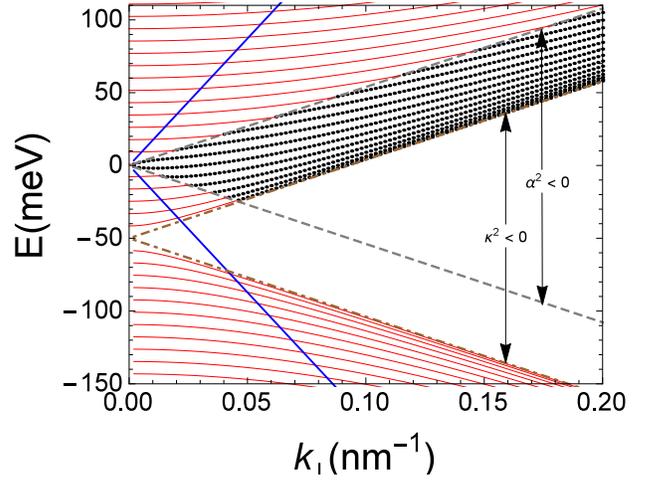}
\caption{(Color online)
Spectrum of confined states (black dotted) in a WSM square QW as a
function of $k_{\perp}$. The spectrum area is limited by gray dashed
and dash-dotted lines of $\alpha=0$ and $\kappa=0$, respectively. Blue
solid lines show the dispersion of electrons at
$\theta=0.1\pi$. Red solid curves show $\pm\epsilon_n$ with
$n=1,2,3,\cdots$ versus $k_{\perp}$ in Eq. (\ref{T-top}) indicating the
resonant positions of incident electrons in areas where $\alpha^2>0$
and $\kappa^2>0$.
}
\label{Fig1:eigen2}
\end{figure}

The discrete energy branches versus the transverse momentum $k_{\perp}=(k_y,k_z)$ are presented by the black dotted curves in Fig. \ref{Fig1:eigen2}.
The understanding of Fig. \ref{Fig1:eigen2}
is \emph{crucial to comprehend} the following results in this work.
The absence of confinement is observed for $k_{\perp}=0$ due to the
\emph{Klein tunneling}, which tells that a massless relativistic
particle can transmit a potential step with unit probability at normal
incidence. At non-normal incidence, the transmission problem for 3D
massless fermions can be represented as a two-dimensional (2D) problem
for massive Dirac fermions, with the effective mass proportional to
the component of $k_{\perp}$ in the extra dimensionality.
The confined states reside in a narrow energy range satisfying
$-k_{\perp}L<\epsilon<k_{\perp}L$ where $\alpha^2<0$ and
$\epsilon>k_{\perp}L-u_0$ where $\kappa^2>0$. The former two
restrictions define the continuum limits of free electrons and holes
propagating along the Weyl QW, whose behavior will be discussed later.
The two lower limits set a turning point of the
energy range at $k_{\perp}L=u_0/2$. Interestingly, branches of confined
state can emerge in very large energy range from the half QW depth
until infinity. They can exist outside
the potential well.
On the continuum edges where $\hat{\alpha}\to 0$,
the energy minima of the branches appear at
\begin{eqnarray}
\epsilon^{\rm min}_n=\frac{n^2\pi^2}{2 u_0}-\frac{u_0}{2}\,,
\end{eqnarray}
for integer $n=1,2,3,\cdots$. As indicated by the black dotted curves
in Fig. \ref{Fig1:eigen2}, these energies are equal to the resonant
energies given by Eq.(\ref{T-top}) for extended electrons on the
continuum edges tunneling through the QW (red curves). The resonance
makes the conversion of confined electrons to free particles become
straightforward and the confinement of electrons in the Weyl QW be
impossible. These phenomena are in striking contrast to the
non-relativistic QW. Nevertheless, Klein tunneling is suppressed away
from the edges and the confinement of electrons in a 1D potential is
allowed \cite{Vas06}. For large $k_{\perp}$,
the dispersion branches of the confined states also coincide with the
resonant transmission conditions given by Eq. (\ref{T-top}).

One of the outstanding properties of Weyl electrons is their immunity
to scattering by potential variations. To understand the physical
mechanism behind it, here we study the propagating
behavior of an electron (hole) incident to the 1D square potential at an angle
$\theta=\textrm{arcsin}(k_{\perp}L/|\epsilon|).$
The transmission coefficient $T\equiv\vert t\vert^2$ and
the reflection coefficient $R=1-T$ can be obtained analytically with
\begin{eqnarray}
t=
\frac{(g_{-} - g_+)(f_- - f_+)e^{-i \alpha} }{(g_{-} - f_-)(g_+ - f_+)
e^{  i \kappa }-(g_{-} - f_+)(g_+ - f_-)e^{  -i \kappa }}\,,
\label{tran2}
\end{eqnarray}
where $f_-=(\beta+i\alpha)/(\epsilon+\gamma)$,
$f_+=(\beta-i\alpha)/(\epsilon+\gamma)$,
$g_{-}=(\beta+i\kappa)/(\epsilon+u_0+\gamma)$, and
$g_+=(\beta-i\kappa)/(\epsilon+u_0+\gamma)$.

\begin{figure}[t!]
\includegraphics[width=8.3cm]{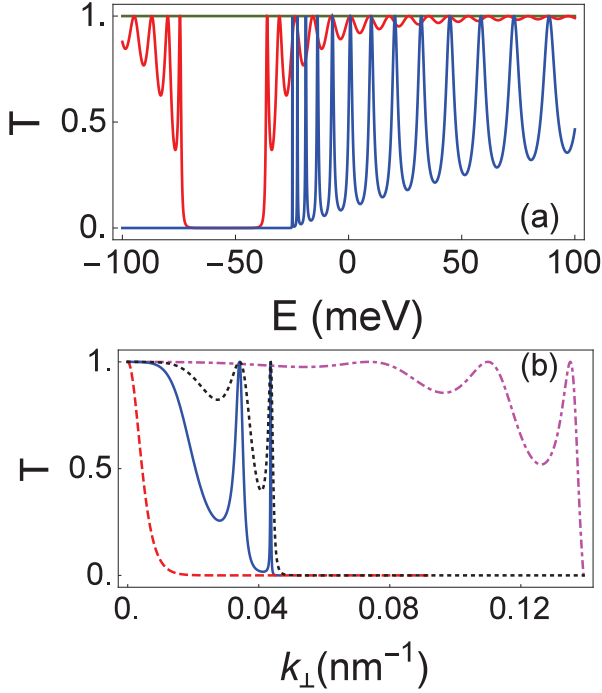}
\caption{(Color online) $T$ as a function of:
(a) $E$ for different $\theta=0$
(green), $0.1\pi$ (red), $0.4\pi$ (blue), and
(b) $k_{\perp}$ for $E=-75$ meV (black, dotted), $-50$ meV
(red, dashed), $-25$ meV (blue, solid), $50$ meV (pink, dash-dotted).
}
\label{Fig2:T-E}
\end{figure}

In Fig. \ref{Fig2:T-E} we present the transmission versus $E$ and $k_{\perp}$.
A careful scrutinization reveals that $T$ shows aperiodic maxima as a function of $E$ when $\kappa$ is real and equals exactly
to an integer multiple of $\pi$, i.e., $\kappa =n\pi$, which gives
\begin{eqnarray}
\epsilon^{\pm}_n = \pm \sqrt{k^2_{\perp}L^2+(n \pi)^2}-u_0.
\label{T-top}
\end{eqnarray}
Here the $+$ ($-$) sign corresponds to values at the upper (lower)
continuum edges.
On the other hand, the reflection coefficient $R$ displays aperiodic
maxima (coinciding with the minima of $T$) as a function of the incident
energy when $\kappa$ is equal to a half integer multiple of $\pi$, i.e.,
$\kappa=(n+1/2)\pi$. Also, the minimal value of $T$ decreases with the
increase of the incident angle or the energy.

On the continuum edges where $\alpha\to 0$ and
$|\epsilon|\to k_{\perp}L$, ideal transmissions occur at the same
energies as those of the confined states on the other side of the
continuum edges, i.e., $\epsilon^+_n=\epsilon^{\rm min}_n$, in the
energy range $\epsilon>u_0/2$, where confined states exist. This can
be understood as coherent interferences of multiple transmissions when
the incident energy resonates with the confined states at any fixed
$k_{\perp}$ as displayed in Fig. \ref{Fig1:eigen2}.

In Fig. \ref{Fig2:T-E}(a) we plot $T$ versus $E$ for
$\theta=0$ (green), $0.1\pi$ (red) and $0.4\pi$ (blue).
The oscillating transmission spectrum is symmetrically distributed with
respect to the energy $\epsilon_0=-2u_0/\cos^2\theta$. A transmission
gap between $\epsilon^-_0=-u_0/(1-\sin\theta)$ and
$\epsilon^+_0=-u_0/(1+\sin\theta)$ appears corresponding to the area
$\alpha^2>0$ and $\kappa^2<0$ in Fig. \ref{Fig1:eigen2}. For large
$\theta$, $T$ can be greatly reduced even when $E$ is much higher than the barrier top,
or lower than the
well bottom.

The dependence of $T$ on
$k_{\perp}$ or $\theta$ at $\epsilon=-1.5u_0$,
$-u_0$, $-0.5u_0$ and $u_0$ is presented in Fig. \ref{Fig2:T-E}(b).
For $\theta\to 0$, $T$ is equal to 1.
The increase of $\theta$ greatly suppresses $T$ except for "magic" angles,
where the condition in Eq. (\ref{T-top}) for resonance is satisfied.
For valence electrons of energies $\epsilon$ less than $-u_0/2$,
only those with $\theta$ lower than critical value can transmit across the QW.
This means also that a QB is more efficient to collimate conduction electron beam along
the normal line than a QW.

Based on the above momentum filtering property of WSM QWs and QBs,
we can design p-n-p or n-p-n transistors made of WSM heterojunctions as collimators of electron beams.
One of the prototypes can be made from two types of WSMs slabs, A and B, with well-matched lattices but different Weyl node energies.
Using MBE technologies, a planar transistor can be fabricated by sandwiching a B slab of thickness in the nano scale between two A slabs.
When an electron beam with electron energies near the Weyl node energy of B passes through the A-B-A nanostructure,
electrons of large transverse momentums will be blocked and filtered out.
The outgoing electron beam becomes then normal to the A-B interfaces.
By connecting a series of parallel transistors with slightly different Weyl node energies in B,
we can make collimators with wider band widths.

\begin{figure}[t!]
\begin{center}
\includegraphics[width=8.7cm]{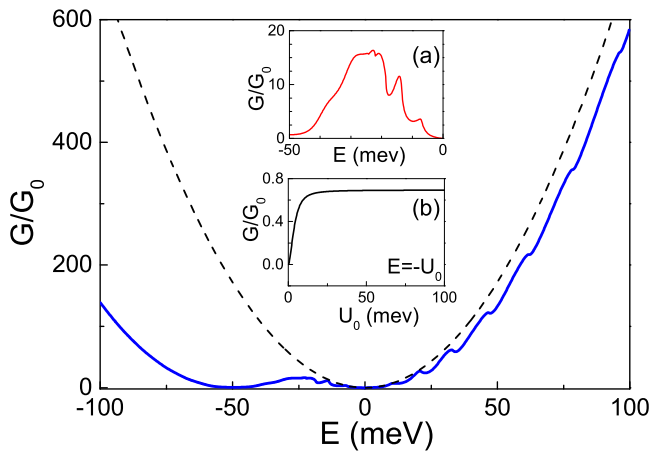}
\end{center}
\caption{(Color online)
$G$ in unit of $G_0$ versus $E$.
$G$ for $U_0=0$ is shown as the dashed line for comparison.
Inset (a) amplifies the part in energy range $-U_0<E<0$.
Inset (b) shows the $G/G_0$ minima at $E=-U_0$ versus $U_0$.
}
\label{Fig3:G_E}
\end{figure}

To this end, we calculate the 3D conductance
$G= G_0 L^2 \int_{0}^{\infty} k_{\perp} d k_{\perp} T(E,k_{\perp})$
with $G_0\equiv S e^2/(2\pi h L^2)$ which
gives the zero-temperature linear conductance of a system with cross
section area $S$ over all angles at the Fermi energy equal to $E$.
$G(E)$ is
presented by the solid curve in Fig.\ref{Fig3:G_E} with part of it
zoomed in inset (a). Its counterpart of free Weyl electrons
$G=G_0 E^2 L^2/[2(\hbar V_F)^2]$ in the case of $U_0=0$ is shown by the
dashed curve for the sake of comparison. In energy range $E > -U_0/2$
where confined states exist in the QW, the conductance is only slightly
lower than that without QW. It displays an oscillating behavior with
the minima at $\epsilon^{\rm min}_n$
corresponding to $E = -24.28$ meV ($n$=1), $-22.13$ meV ($n$=2),
$-18.55$ meV ($n$=3), $-13.53$ meV ($n$=4), and $-7.08$ meV ($n$=5).
In contrast, in energy range $E<-U_0/2$, the conductance is much
smaller than that in the absence of QW and the spectrum curve becomes
smooth. This conductance drop originates from the blocking of electrons
with large incident angle towards the QW as explained in Fig.
\ref{Fig1:eigen2}, and the transmission gap shown in Fig.
\ref{Fig2:T-E}(a). Specifically, at energy $E=-U_0$, another
conductance minimum appears when only electrons propagating almost normally
to the QW plane can pass across the junction with a transmission
coefficient
\begin{eqnarray}
T^{\rm min}=\frac{2(u_0^2-k_{\perp}^2 L^2)}{u_0^2+u_0^2
\cosh(2 k_{\perp} L)-2 k_{\perp}^2 L^2 }.
\label{minimal_G}
\end{eqnarray}
The dependence of the minimal conductance on $U_0$ is shown in the
inset of Fig. \ref{Fig3:G_E}(b). The minimal conductance at $E=-U_0$
increases with $U_0$ and saturates at a value about $G/G_0=0.69$ when
$U_0>20$ meV. In the absence of the potential or when $U_0$ approaches
zero, it merges with the zero conductance at the Weyl node energy $E=0$.

In summary, we have shown that confined states are allowed for finite transverse
wavevector in some energy range. The confined states even emerge at
the edges of the continuum of free particles, and this characteristic
brings about a great impact on the transmission. We find that the
electronic transmission through the nanostructures is dependent on the
incident angles. The Klein tunneling mechanism is generally suppressed
for the obliquely incident electrons which can be confined by
electrostatic potentials. However, in addition to the full transmission
for normally incident electron beam, there are finite ¡°magic angles¡±
of the incidence for full transmission, which are associated to
Fabry-P\'{e}rot resonances at both interfaces. Specifically, for
electrons have energy equal to the QW bottom or to the QB top energy,
only the normal incident electrons can pass the structure and an extra
conductance minimum appears at the energy. The transmission probability
through heterojunctions could be testified by the bulk resistance
measurements.

Our study suggests
that for p-type (n-type) Weyl materials one can apply a QW (QB)
potential to make a p-n-p (n-p-n) transistor and manipulate the
direction and magnitude of the charge current. When the bottom (top)
energy of the QW (QB) is aligned to the Fermi energy, i.e.,
$E_F=-U_0$ ($E_F=U_0$), the charge current can be well collimated with
the linear conductance greatly reduced by the potential. Since the
energy band and electron properties are well protected by unbroken
symmetry in Weyl materials, the corresponding device is expected more
robust than those of normal semiconductors. In this case, n-p-n or
p-n-p nanostructure transistors can be designed to collimate electron
beams. Also, this transistor may be used as a tunable well for
electrostatic quantum confinement, which can be deployed as a single
Weyl fermion pump.

\acknowledgments
W.-L.Y. acknowledges the helpful discussion with H. Jiang and
support by the Natural Science Foundation of Jiangsu Province of China
under Grant No. BK20141190 and the NSFC under Grant Nos. 11474211 and 11204197.
X.-F.W. appreciates support from NSFC under Grant Nos. 11074182
and 91121021.
A.M.O. kindly acknowledges support by Narodowe Centrum Nauki
(NCN, National Science Center) under Project No. 2012/04/A/ST3/00331.

\end{document}